\documentclass[paper,notoc,nohyper]{JHEP}

\usepackage{amssymb,epsfig,eucal,amsmath,cite} \psfull

\def\cF{{\mathcal{F}}}

\def\cO#1{{\CMcal O}\left( {#1} \right)}
\def\asb{{\bar \alpha}_{\mbox{\scriptsize s}}}
\def\as{\alpha_{\mbox{\scriptsize s}}}
\def\ga{\gamma}
\def\om{\omega}

\def\half{\mbox{\small $\frac{1}{2}$}}

\def\ho{\half\om}

\def\eff{\mathrm{eff}}
\newcommand{\op}{\om_{\mathbb{P}}}           
\newcommand{\gb}{\bar{\gamma}}

\def\cFL{\CMcal A}
\def\cFH{\CMcal B}
\def\cFl{a}

\def\tbar{{\bar t}}
\def\epj#1#2#3{{ \it Eur. Phys. J. } {\bf {C#1}} (#2) #3}
\newcommand{\Ai}{{\rm Ai}}                     
\newcommand{\ab}{\bar{\al}_\mathrm{s}}
\newcommand{\al}{\alpha}
\newcommand{\Bi}{{\rm Bi}}                  

\newcommand{\Dt}{{\mathcal D}_{t}}     
\renewcommand{\d}{\delta}

\newcommand{\de}{\partial}
\newcommand{\dif}{{\rm d}}
\newcommand{\difg}{{\dif\ga\over2\pi\ui}}
\newcommand{\difo}{{\dif\om\over2\pi\ui}}

\newcommand{\esp}[1]{{\rm e}^{#1}}
\newcommand{\F}{{\CMcal F}}                  

\newcommand{\G}{{\CMcal G}}                  

\newcommand{\GG}{{\mathcal G}}     

\newcommand{\K}{{\CMcal K}}                    
\newcommand{\KK}{{\mathcal K}}                

\newcommand{\kk}{{\boldsymbol k}}

\newcommand{\kw}{{\kappa}}                     
\newcommand{\La}{\Lambda}

\newcommand{\omb}{\bar{\om}}                 
\newcommand{\ord}{{\CMcal O}}                

\newcommand{\si}{\sigma}
\newcommand{\Th}{\Theta}
\newcommand{\tb}{\bar{t}}                          
\newcommand{\ts}{\textstyle}
\newcommand{\ui}{{\rm i}}
\newcommand{\xib}{\bar{\xi}}                     
\title{A collinear model for small-$\boldsymbol x$ physics.
\thanks{Work
    supported by E.U. QCDNET contract FMRX-CT98-0194.}}

\author{M. Ciafaloni,$^1$ D. Colferai$^1$ and G.P. Salam$^2$\\
  $^1$ Dipartimento di Fisica, Universit\`a di Firenze and INFN,
  Sezione di Firenze\\
  Largo E. Fermi 2, 50125 Firenze, Italy\\
  $^2$ INFN, Sezione di
  Milano, Via Celoria 16, 20133 Milano, Italy\\
  E-mail: \email{Ciafaloni@fi.infn.it, Colferai@fi.infn.it,
    Gavin.Salam@mi.infn.it}} 
  
\abstract{We propose a simple model for studying small-$x$ physics in
which we take only the collinearly enhanced part of leading and
subleading kernels, for all possible transverse momentum
orderings. The small-$x$ equation reduces to a second order
differential equation in $t\equiv\log k^2/\La^2$ space, whose
perturbative and strong-coupling features are investigated both
analytically and numerically. For two-scale processes, we clarify the
transition mechanism between the perturbative, non Regge regime and
the strong-coupling Pomeron behavior.}

\preprint{hep-ph/9907nnn\\Bicocca--FT--99--23\\
	Firenze--DFF--342--7--99\\July 1999}

\keywords{QCD, NLO computations}

\begin{document}
\section{Introduction}
\label{sec:intro}

The problem of subleading logs corrections to the BFKL
equation~\cite{BFKL,NLL,FL,CC98} has been recently investigated by
several authors~\cite{BV,Ross,Levin,Th99,Sa98,CiCo98b,CCS1} in
connection with the small-$x$ behavior of structure functions and of
two-scale processes~\cite{hera,2dis,fjet}.

We have advocated~\cite{Sa98,CiCo98b,CCS1} in this context, the
importance of incorporating Renormalisation Group (RG) improvements
and collinear physics in a new form of the small-$x$ equation. The
latter has the virtue of including all collinearly enhanced higher
order contributions to the kernel, and provides stable estimates of
the resummed gluon anomalous dimension and of the hard Pomeron.

Here we present a simple, but powerful tool for studying the problem
of small-$x$ physics that we call {\em the collinear model} --- namely
a model where all and only the collinearly-enhanced physics is
correctly included, in particular the full dependence on the one-loop
running coupling, the splitting functions, and the so-called energy
scale terms~\cite{CC98,Sa98}.

The model has the properties of correctly reproducing the one-loop
renormalisation group results and of being symmetric (given a problem
with two transverse scales, its results do not depend on which of the
scales is larger). These are properties desired from the resummation
of the NLL corrections in the case of the full BFKL kernel.

While it does not correctly resum the series of leading and subleading
logarithms of $s$ (i.e.\ the non-collinear part of the problem), in
that region it does have a structure which qualitatively is very
similar to that of the BFKL equation, and can usefully serve as a
model. In contrast to the BFKL equation, it is very easily soluble, as 
a Schr\"odinger-like problem.


\section{The collinear model}\label{s:cm}

Let us start recalling~\cite{CC98} that, in order to specify a small-$x$
model for a hard process with two scales $k$ and $k_0$, one should
specify the energy scale $s_0$, the kernel and the impact
factors. They all enter the $\kk$-factorised form of the cross section
\begin{equation}
 \si_{AB}(Y;k,k_0)=\int\difo\left(\frac{s}{s_0(k,k_0)}\right)^\om h_A(k)
 \,\G_\om(k,k_0)\,h_B(k_0)\,,
 \label{fatt}
\end{equation}
where $Y\equiv\log s/s_0$ and
\begin{equation}
 \G_\om=[\om-\K_\om]^{-1}	\label{defg}
\end{equation}
is the gluon Green's function. In the following, we refer normally to
the symmetrical factorised scale $s_0=kk_0$, by keeping in mind that
for $k\gg k_0$ (or viceversa), one can switch to the relevant Bjorken
variable $k^2/s$ (or $k_0^2/s$) by the similarity transformation
induced by $(k/k_0)^\om$ on the kernel and the Green's function.

\subsection{Definition of the model}

Our collinear model is then defined --- in logarithmic variables
$t\equiv\log k^2/\La^2$ --- by specifying the kernel
$\K_\om(\as(t),t,t')$ whose $\as(t)$ expansion corresponds to higher
and higher order collinear singular kernel for both $k\gg k'$ and
$k'\gg k$.
\FIGURE{\epsfig{file=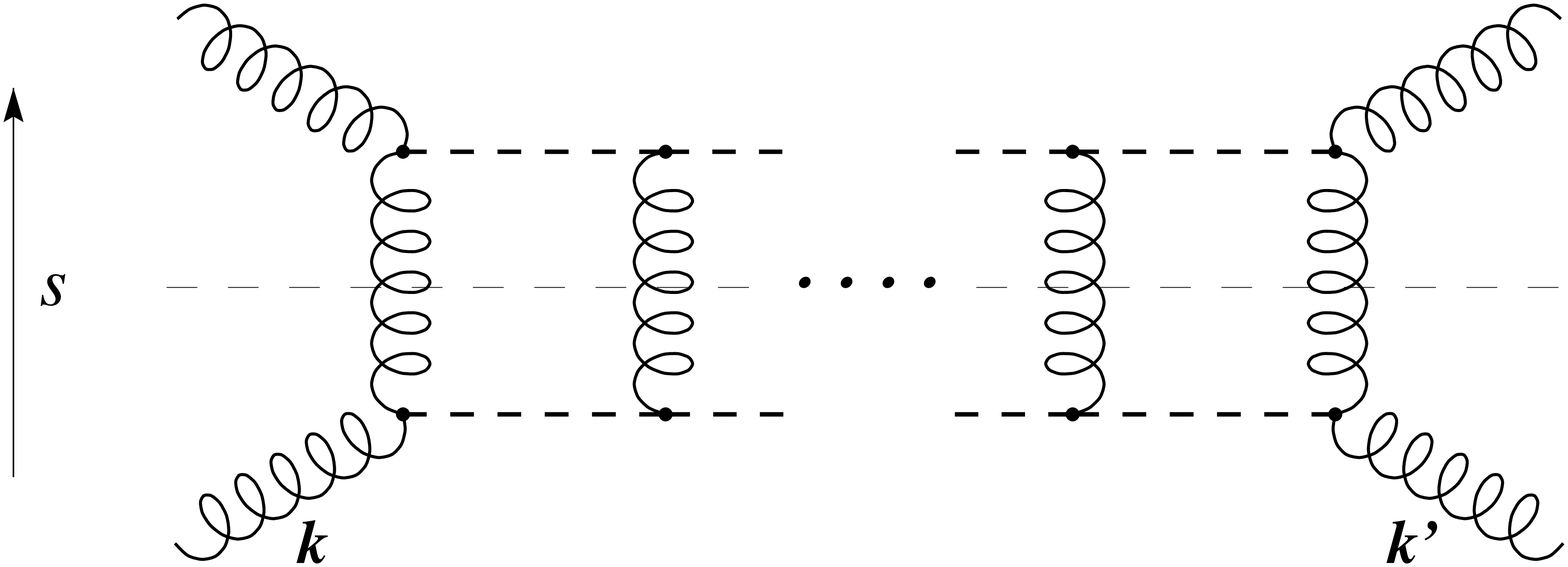,width=0.6\textwidth}
    \caption{Subleading kernels being resummed by $\K_\om(\kk,\kk')$
	in Eq.~(\ref{eq:kernel}). Wavy (dashed) lines denote
	high-energy (low-energy) gluon exchanges, corresponding to
	the $\ab/\om$ ($A_1\ab$) part of the gluon anomalous
	dimension.}
    \label{f:nucleo}}

In other words, our model is such that it reproduces the DGLAP
limits~\cite{DGLAP} for branchings with ordered transverse momenta,
and the anti-DGLAP limit for branchings with anti-ordered transverse
momenta. Accordingly, the kernel governing a small-$x$ branching of a
gluon with transverse momentum $k'$ into a gluon with transverse
momentum $k$ is
(Fig.~\ref{f:nucleo})
\begin{align}
  \label{eq:kernel} 
 {1\over\om}\KK_\om(t,t')&\equiv{kk'\over\om}\K_\om(k,k')\\
  &={\asb(t)\over\om}\exp\left\{-{1+\om\over2}
  (t-t')+A_1(\om)\int_{t'}^t\ab(\tau)\,\dif\tau\right\}\Th(t-t')
  +(t\leftrightarrow t')\,.\nonumber
\end{align}
The term proportional to $A_1(\om)$ comes from the summation of
the non-singular part of the DGLAP splitting function, including
a full treatment of the running coupling:
\begin{equation}
 \ga_{gg}(\ab,\om)-{\ab\over\om}=A_1(\om)\ab(t)+\ord(\ab^2)\,,
 \qquad\ab\equiv{N_c\as\over\pi}={1\over bt}\,.
\end{equation}
Therefore the kernel for a single small-$x$ branching actually resums
many branchings, of which the last (and only the last) is governed by
the $1/\om$ part of the splitting function.

The scale of $\asb$ is the larger of the two scales involved in the
branching, as implied by DGLAP. Finally, the factor $kk'$ is due to the 
fact that we work in $t$-space instead of $\kk$-space, and the
remaining one $\exp[-\ho(t-t')]$ is due to the choice of scaling
variable $kk_0/s$, as explained before.

The collinear properties of the kernel (\ref{eq:kernel}) can also be
seen in $\ga$-space, by the expansion
\begin{equation}
  \K_{\om}(k,k')=\sum_{n=0}^{\infty}[\ab(k^2)]^{n+1}\;
 K_n^{\om}(k,k')\,,           \label{serie}
\end{equation}  
where the scale-invariant kernels $K_n^\om$ have eigenfunctions
$\esp{(\ga-\half)t}$ and eigenvalue functions $\chi_n^\om(\ga)$ given
by~\cite{CiCo98b}
\begin{equation}
 \chi^{\om}_n(\ga)={1\cdot A_1(A_1+b)\cdots(A_1+(n-1)b)\over
 (\ga+\ts{1\over2}\om)^{n+1}}+{1\cdot(A_1-b)(A_1-2b)\cdots(A_1-nb)\over
 (1-\ga+\ts{1\over2}\om)^{n+1}}\,.\label{chin} 
\end{equation}

Of particular interest is the NL expansion
\begin{align} 
 \int\dif t'\;\KK_\om(t,t')\esp{-(\ga-\half)(t-t')}&=
 \left(\ab(t)\chi_0^\om+\ab(t)^2\chi_1^\om\right)\label{nlesp}\\
 &\simeq\ab(t)\left[\chi_0^0-\frac12\om\left(\frac1{\ga^2}+
 \frac1{(1-\ga)^2}\right)\right]+\ab(t)^2\chi_1^0+{\rm NNL}\nonumber
\end{align}   
whose $\om$-dependence can be reabsorbed~\cite{CCS1} in a redefinition
of impact factors and kernel, to yield the leading and next-to-leading
BFKL kernels of the model:
\begin{subequations}
  \begin{align}
    \chi_0 &= \frac1\ga  +  \frac1{1-\ga}\,, \\
    \chi_1 &= \frac{A_1}{\ga^2}  +  \frac{A_1-b}{(1-\ga)^2} 
               - \frac{\chi_0}{2}\left(\frac{1}{\ga^2} 
                 + \frac{1}{(1-\ga)^2}\right)\,,
  \end{align}
\end{subequations}
where the renormalisation scale for $\asb$ has been taken as $t$,
i.e.
$$
\KK_{\rm BFKL}(\as(t),\ga) = \asb(t) \chi_0(\ga) + \asb^2\chi_1(\ga)\,.
$$
At leading order, this reproduces the poles at $\ga=0$ and $1$ of the
true BFKL kernel. At next-to-leading order it reproduces the $\ga=0,1$
quadratic and cubic poles of the true BFKL kernel. Numerically the
leading order kernel differs from BFKL quite significantly
numerically, but retains a very similar structure --- a saddle point
at $\ga=1/2$, implying a power growth of the cross section, and
diffusion. The double and triple-polar part of the next-to-leading
kernel turns out to be very close, even numerically, to the full BFKL
NLO kernel, reproducing it to better than $7\%$ over the whole range
of $\ga$ from $0$ to $1$ (note though that this collinear kernel has
single polar and other parts, so that as a whole it may not be quite
this close to the full BFKL NLO kernel). This suggests that
collinearly enhanced effects dominate the NLO kernel.

\subsection{First order formulation}

The main advantage of our collinear kernel from the point of view of
this article is its relative simplicity. Specifically it can be
written in factorised form:
\begin{equation}
  \label{eq:kernelfacts}
 \KK_\om(t,t') = U(t) \,V(t')\, \Theta(t-t') + U(t') \,V(t)\, \Theta(t'-t)\,,
\end{equation}
where
\begin{subequations}
\begin{align}
 U(t) &=\asb(t)\exp\left\{-{1+\om\over2}t+A_1(\om)
  \int^t\ab(\tau)\,\dif\tau\right\}\,,\\
 V(t) &=\exp\left\{{1+\om\over2}t-A_1(\om)\int^t\ab(\tau)
 \,\dif\tau\right\}\,.
\end{align}
\end{subequations}
This allows us to recast the homogeneous BFKL equation\footnote{%
Since we work in $t$-space, the density $\cF(t)$ differs by a factor
of $k$ from the customary~\cite{CCS1} BFKL solution $\F(k)$, at scale
$kk_0$.},
\begin{equation}
  \label{eq:homogen}
 \om\cF(t)\equiv\om k\F_\om(k) = U(t) \int_{-\infty}^t\dif t'\;V(t') \,\cF(t') + 
           V(t) \int^{ \infty}_t\dif t'\;U(t') \,\cF(t')
\end{equation}
as a differential equation. Dividing $\cF$ into two parts,
\begin{subequations}\label{eq:ABFdef}
  \begin{align}\label{eq:Adef}
    \cFL(t) &= U(t) \int_{-\infty}^t\dif t'\;V(t') \,\cF(t')\\
    \cFH(t) &=   V(t) \int^{ \infty}_t\dif t'\;U(t') \,\cF(t')\label{eq:Bdef}\\
    \om\cF(t) &= \cFL(t) + \cFH(t)
  \end{align}
\end{subequations}
and taking the derivative leads to a pair of coupled differential
equations: 
\begin{subequations}
  \label{eq:DEUV}
  \begin{align}
    \frac{\dif\cFL}{\dif t} &= \frac{U'}{U} \cFL + U V \cF\,,\\
    \frac{\dif\cFH}{\dif t} &= \frac{V'}{V} \cFH - U V \cF\,.
  \end{align}
\end{subequations}
For the specific kernel \eqref{eq:kernel} that we consider, we have 
\begin{subequations}
\label{eq:DE}
\begin{align}
  \frac{\dif\cFL}{\dif t} &= \left(-\frac{1+\om}{2} +
    A_1\asb + \frac{\asb'}{\asb}\right)\cFL +\asb\cF \\
 \frac{\dif\cFH}{\dif t} &=\left(\frac{1+\om}{2}-A_1\asb\right)\cFH-\asb\cF
\end{align}
\end{subequations}
Since we have two coupled equations, there are two independent
solutions. Examining the equation for large and positive $t$, where
$\asb$ is small, one sees that they can be classified as a regular
solution
\begin{equation}
  \label{eq:FRbigt}
  \cF_R \sim \exp\left(-\frac{1+\om}{2}t\right) ,
\end{equation}
which is dominated by $\cFL$, and an irregular solution 
\begin{equation}
  \label{eq:FIbigt}
  \cF_I \sim \exp\left(\frac{1+\om}{2}t\right) ,
\end{equation}
dominated by $\cFH$. 

\subsection{Second order formulation}

The coupled set of differential equations \eqref{eq:DEUV} can be
recast in the form of a simple second order equation for $\cF$. In
fact, by using \eqref{eq:ABFdef}, we can first rewrite \eqref{eq:DEUV}
in the form
\begin{equation}\label{eq:2ndorda}
  \om \cF = \left[ \left(\partial_t - \frac{U'}{U}\right)^{-1}
                -\left(\partial_t - \frac{V'}{V}\right)^{-1}\right]
  UV \cF\,.
\end{equation}
Then, in order to eliminate the resolvents appearing in
\eqref{eq:2ndorda} we introduce the operator
\begin{equation}
  \label{eq:oper}
  \Dt = \left(\partial_t + \frac{U'}{U} - \frac{w'}{w}\right)
            \left(\partial_t - \frac{U'}{U}\right)
      = \left(\partial_t + \frac{V'}{V} - \frac{w'}{w}\right)
            \left(\partial_t - \frac{V'}{V}\right),
\end{equation}
and the wronskian
\begin{equation}
  \label{eq:wronskian}
  w(t) =W[U,V]\equiv UV'-U'V=\asb(t)\left(1 + \om - 2A_1\asb -
    \frac{\asb'}{\asb} \right)\,.
\end{equation}
By applying $\Dt$ to (\ref{eq:2ndorda}) we finally obtain
\begin{equation}
  \label{eq:2ndord}
  \om\Dt\cF=\left(\frac{U'}{U}-\frac{V'}{V}\right)UV\cF=-w(t)\cF,
\end{equation}
which is the second order formulation that we were looking for.

Equation \eqref{eq:2ndord} can be recast in normal form by the
similarity transformation
\begin{equation}
  \label{eq:simtrans}
  \cF ={\rm const}\cdot\sqrt{w(t)}\,h(t)\,,
\end{equation}
which leads to a Schr\"odinger type equation,
\begin{align}\label{eq:schr}
  (-\partial_t^2 + V_{\eff})\, h &= 0\\
  V_{\eff} &=\frac{1}{4}\left(\frac{w'}{w}\right)^2-\frac12
 \left(\frac{w'}{w}\right)'-\frac{w}{\om}+\frac{U''V'-V''U'}{w}\,.\label{veff}
\end{align}
Note that the above derivation is valid for any form of the running
coupling $\asb(t)$ which extrapolates the perturbative form
$(bt)^{-1}$ into the strong coupling region around the Landau pole
$t=0$.
\FIGURE{%
   \epsfig{file=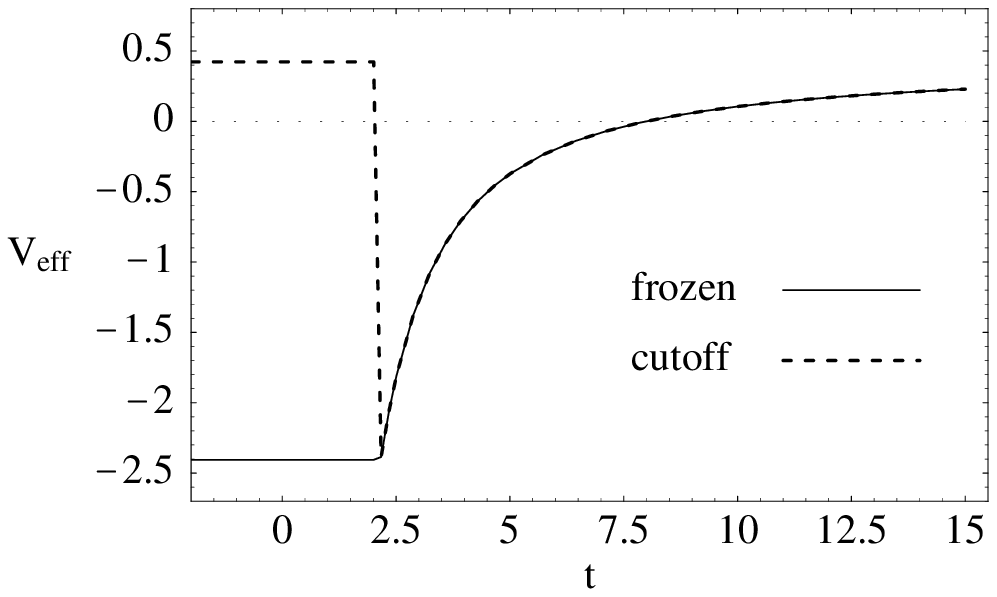,width=0.7\textwidth}
    \caption{Qualitative $t$-dependence of the effective potential for
	the regularisations of type (a) and (b) of the coupling strength.}
    \label{f:potenziale}}

In what follows we consider various regularisations of the Landau pole,
in particular:
\begin{align}
  \mbox{(a)} \qquad \asb(t) &=\frac{1}{bt} \Theta (t - \tbar),
  \qquad\qquad &\mbox{(cutoff case)}, \label{eq:ascut}\\
  \mbox{(b)} \qquad \asb(t) &=\frac{1}{bt} \Theta (t - \tbar)
    + \frac{1}{b\tbar} \Theta ( \tbar - t),
  \qquad\qquad &\mbox{(frozen $\asb$ case)} \label{eq:asfrz},
\end{align}
where $\tbar>0$ sets the boundary of the perturbative behavior.

It is obvious that such different forms will change the form of the
potential  (Fig.~\ref{f:potenziale}) and thus the boundary
conditions on $\cF$ (or $h$) coming from the strong  coupling region. 
A similar Schr\"odinger formulation was found~\cite{CC96} for the
(Airy) diffusion model~\cite{GLR,CoKw89} with running coupling, with a 
potential which roughly corresponds to the bottom of the well in
Fig.~\ref{f:potenziale}.

We consider in particular the solution $\cF_R(t)$ ($\cF_L(t)$) of the
homogeneous equation \eqref{eq:2ndord} which is regular for $t \to
+\infty$ ($t\to-\infty$). If both conditions are satisfied, $\cF_R=\cF_L$
is an eigenfunction of the BFKL equation. The pomeron singularity
$\om=\op$ is the maximum value of $\om$ for which this occurs (ground
state). 

If $\om > \op$, $\cF_R\neq\cF_L$ and the two solutions have rather
different features. Due to the locality of the differential equation,
$\cF_R$ is \emph{independent} of the regularisation in the region
$t>\tbar$. On the other hand, $\cF_L$ will be dependent on the
behavior of $\asb(t)$ for $t<\tbar$. For instance, in the case of
$\asb(t)$ being frozen \eqref{eq:asfrz}, $\cF_L$ has the exponential
behavior
\begin{equation}
  \cF_L(t)\sim\esp{\kappa t}\,,\qquad(t<\tb)\,,
\end{equation}
where $\kappa=\sqrt{V_\eff(t<\tb)}$ is found from \eqref{veff} to be
\begin{equation}
  \kappa^2 = \left[ \frac12(1+\om) - A_1\asb(\tbar)\right]
             \left[ \frac12(1+\om) - A_1\asb(\tbar) -
             \frac{2\asb(\tbar)}{\om}\right].	\label{kapa}
\end{equation}
On the other hand, if $\asb(t)$ is cutoff, it is simpler to use
directly the first-order formulation, \eqref{eq:DEUV}. We note that we 
can write
\begin{equation}
  \om\cF_L= \cFL_L + \cFH_L= \asb(t) \cFl_L(t) + \cFH_L(t)\,,
\end{equation}
where $\cFl_L$ and $\cFH_L$ are {\em continuous} at $t=\tbar$,
and are exponentially behaved, $\cFl_L \sim \cFH_L \sim
\exp(\half(1+\om)t)$, for $t<\tbar$. By using \eqref{eq:DE} we then
obtain
\begin{equation}
  \left.\frac{\cFL_L}{\asb(t)}\right|_{t=\tbar_+} = 
   \cFl_L(\tbar) = \frac{\cFH_L(\tbar)}{\om(1+\om)},
\end{equation}
which defines the boundary condition for the $\cFL_L$ and $\cFH_L$
components of $\cF_L$ in this case.

\subsection{Factorisation of non-perturbative effects}
\label{sec:fact}

The basic tool for describing BFKL evolution is the Green's function
$\GG_\om(t,t_0)\equiv kk_0\G_\om(k,k_0)$, which satisfies the
inhomogeneous small-$x$ equation
\begin{equation}
  \label{eq:green}
  \om \GG_\om(t,t_0)=\delta(t-t_0)+\KK_\om\otimes\GG_\om(t,t_0),
\end{equation}
and is supposed to be well-behaved for $t,t_0\to\pm \infty$. The
problem of factorisation is the question of the (in)dependence on the
non-perturbative strong-coupling region.

For $t\neq t_0$, $\GG_\om$ satisfies the same differential equation as
$\cF$ and is thus a superposition of two independent solutions. The
large-$t$ behavior implies a regularity condition and suggests the
expression
\begin{equation}\label{eq:fact}
  \GG_\om(t,t_0) = \cF_R(t)\cF_L(t_0)\Th(t-t_0)+\cF_L(t)\cF_R(t_0)
 \Th(t_0-t)\,,\qquad (t\neq t_0)
\end{equation}
where $\cF_R$ ($\cF_L$) is the regular solution for $t\to+\infty$
($t\to-\infty$) defined in the previous subsection.

Actually \eqref{eq:fact} is a rigorous consequence of the second-order
formulation. In fact, $\GG_\om$ satisfies the differential equation
\begin{equation}
 \left(\Dt+{w\over\om}\right)\GG_\om={1\over\om}\Dt\d(t-t_0)\,.
 \label{edg}
\end{equation}
With a little thought, one can realize that $\GG_\om$ must contain a
delta function term in the form
\begin{equation}
 \GG_\om(t,t_0)={1\over\om}\d(t-t_0)+\hat{\GG}_\om(t,t_0)\label{scomp}
\end{equation}
and hence
\begin{equation}
 \left(\Dt+{w\over\om}\right)\hat{\GG}_\om=-{w\over\om^2}\d(t-t_0)
 \label{edhg}
\end{equation}
showing that $\hat{\GG}$ is continuous function at $t=t_0$ with
discontinuous derivative. Eq.~(\ref{edhg}) is an inhomogeneous
Schr\"odinger type equation with a delta source, and its solution can
be found in standard textbooks to be just like the RHS of
Eq.~(\ref{eq:fact}) also for $t=t_0$. We conclude that
\begin{equation}
  \GG(t,t_0)={1\over\om}\d(t-t_0)+\cF_R(t)\cF_L(t_0)\Th(t-t_0)+
 \cF_L(t)\cF_R(t_0)\Th(t_0-t)\label{gdelta}
\end{equation}
with the normalisation
\begin{equation}
  \cF_{R,L}(t)=\frac{\sqrt{w(t)}}{\om}h_{R,L}(t),\qquad\qquad
  W[h_R,h_L]\equiv h_R h'_L - h_R'h_L = 1\,.	\label{fh}
\end{equation}
The main consequence of \eqref{eq:fact} is that the regularisation
dependence is factorised away in $\cF_L$, whenever $t$ or $t_0$ are
large enough. This happens in particular in the collinear limit
$t-t_0\gg 1$ relevant for structure functions. 

\section{Solutions: analytical features}\label{s:af}

The collinear model just defined can be solved in principle as a
Schr\"odinger problem by known analytical and numerical techniques and 
for both types of solutions occurring in the Green's function
(\ref{defg}) (i.e., the left-regular and the right-regular ones).

The regular solution $\cF_R$ is, for large $t$, perturbative,
i.e., independent of the potential in the strong coupling region
$t\leq\tb$, while the left-regular one $\cF_L$ is dependent on the
strong coupling boundary conditions through a reflection coefficient
$S(\om)$ of the $S$-matrix, which occurs in its expression for $t>\tb$:
\begin{equation}
 \cF_L(t)=\cF_I(t)+S(\om)\cF_R(t)\,,	\label{fleft}
\end{equation}
where $\cF_R$ ($\cF_I$) denote the regular (irregular) solution for
$t>\tb$, with the normalisation $W[h_R,h_I]=1$.

This allows to rewrite the Green's function for $t>t_0$ and
$\om>\op$ in the form
\begin{equation}
 \GG_\om(t,t_0)=\cF_R(t)\left[\cF_I(t_0)+S(\om)\cF_R(t_0)\right]\,.\label{fattg}
\end{equation}
If both $t,t_0\gg1$, but $t-t_0=\ord(1)$, Eq.~(\ref{fattg}) is
dominated by the first term, the second being suppressed exponentially 
in $t_0$. This term is, on the other hand, defined by boundary
conditions for $t,t_0\to+\infty$ only, and is therefore independent of 
the strong coupling region. Its analytical and numerical form will be
discussed in more detail in the following.

The second term in Eq.~(\ref{fattg}) carries the regularisation
dependence and contains the leading $\om$-singularities, in particular 
$\op$ (Ref.~\cite{CC96}). In $Y\equiv\log (s/kk_0)$ space, the sum in
Eq.~(\ref{fattg}) defines two asymptotic regimes, as we shall see.

\subsection{Perturbative regime: $\boldsymbol\om$-expansion
	   and WKB limit}

Approximate solutions in the large-$t$ region can be found, as in the
full small-$x$ equation, by the method of the $\ga$-representation and 
$\om$-expansion. The regular solution is approximated by
the expression:
\begin{equation}
  \cF_R(t)\simeq \int\difg\;\exp\left[(\ga-\half)t -
    \frac{X^\om}{b\om}\right]\,,	\label{rapintf}
\end{equation}
with 
\begin{equation}
  \de_\ga X^\om \equiv \chi(\ga,\om) = \chi^\om_0(\ga) + \om
  \frac{\chi^\om_0(\ga)}{\chi^\om_1(\ga)}\,,	\label{XNL}
\end{equation}
where, for the collinear model,
\begin{subequations}\label{chicol}
  \begin{align}
    \chi^\om_0(\ga) &= \frac1{\ga + \half \om}  +  \frac1{1-\ga +
      \half\om},\\
    \chi^\om_1(\ga) &= \frac{A_1}{(\ga + \half \om)^2}  +  \frac{A_1 -
      b}{(1-\ga + \half\om)^2}.
  \end{align}
\end{subequations}
In \cite{CCS1} this representation was extensively studied, and shown
to be a solution of the problem up to next-to-leading order and to all
orders for the collinear structure. But this was only guaranteed to
work true for small values of $\om$ (whereas for the continuation with
the DGLAP anomalous dimensions it is useful to be able to access high
$\om$ as well). Also there was no way of determining the coefficient
of any higher-order error introduced by the procedure.  A comparison
of this representation with the exact solution, as is possible in the
collinear model, is therefore important (cf. Sec.~\ref{sec:RegSoln}).

The expressions (\ref{rapintf}--\ref{chicol}) can be further
specialized in the large-$t$ limit, where (\ref{rapintf}) is dominated 
by a saddle point at $\ga=\gb$:
\begin{equation}
 b\om t=\chi(\gb,\om)\,.	\label{ps}
\end{equation}
The latter is related to the WKB approximation for solving the
differential equation (\ref{eq:schr}). In fact, on the basis of
Eqs.~(\ref{eq:schr}) and (\ref{fh}), one can prove the asymptotic expansion
\begin{equation}
 \cF_R(t)={1\over\om}\sqrt{w(t)\over2\,\kw(t)}\exp\left\{-\int^t\kw(\tau)
 \,\dif\tau\right\}\times\left[1+\ord\left({1\over t}\right)\right]\,,\label{wkb}
\end{equation}
where $\kw(t)$ is defined in terms of the effective potential
(\ref{veff}) as
\begin{equation}
 \kw^2(t)\equiv V_\eff(t)=\frac12\left(1+\om -{2A_1-b\over bt}\right)
             \left[\frac12\left(1+\om-{2A_1-b\over bt}\right) -
             \frac{2}{b\om t}\right]+\ord\left({1\over t^2}\right) \label{Kwkb}
\end{equation}
and is related to the saddle point value $\gb(\om,t)$ of
Eq.~(\ref{ps}) by
\begin{equation}
 \gb=\frac12\left(1+\om+{1\over t}\right)-\kw(t)+\ord\left({1\over t^2}
 \right)	\label{gb}\,.
\end{equation}

It is interesting to note that in the collinear model, because of the
differential equation (\ref{eq:schr}), the present method yields the
irregular solution also, which is obtained by just changing the sign
of the WKB momentum $\kw(t)$, i.e.,
\begin{equation}
 \cF_I(t)={1\over\om}\sqrt{w(t)\over2\,\kw(t)}\exp\left\{
 +\int^t\kw(\tau)\,\dif\tau\right\}\,.	\label{wkbirr}
\end{equation}
This is useful for evaluating the Green's function which according to
Eqs.~(\ref{fattg}), (\ref{wkb}) and (\ref{wkbirr}) takes the form
\begin{align} 
 \GG_{\rm WKB}(t,t_0)={1\over2\om^2}\sqrt{w(t)w(t_0)\over
 \kw(t)\kw(t_0)}&\left[\exp\left\{-\int_{t_0}^t\kw(\tau)\,\dif\tau\right\}
 \right.\label{gwkb}\\
 &\left.+S(\om)\exp\left\{-\left(\int_{t_s}^{t_0}+\int_{t_s}^t\right)
 \kw(\tau)\,\dif\tau\right\}\right]\quad,\quad(t>t_0)\,,\nonumber
\end{align}
where $t_s$ is the zero of $\kw$ and $S(\om)$ has to be determined
from the boundary conditions on $\cF_L$ at $t=\tb$.

\subsection{Strong-coupling features}

For intermediate, small, and negative $t$'s, the collinear model enters 
a second regime, where both kinds of solutions oscillate. Given the
definitions of the effective potential in (\ref{veff}), the regime's
boundary is roughly given by $t<\chi_m/b\om$, where $\kw(t)$ becomes
negative, and $\chi_m$ is the minimum (in $\ga$) of the effective eigenvalue
function $\chi(\ga,\om)$ defined in Eq.~(\ref{XNL}).

A basic question concerning this regime is the spectrum of $\K_\om$,
which provides the $\om$-singularities of the Green's function, which
in turn determine the high-energy behavior of the cross section.

The leading $\om$-singularity is the Pomeron $\op$, i.e., the maximum
$\om$ value for which $\cF_L$ and $\cF_R$ match each other, providing a
zero energy bound state in the potential (\ref{veff}).
The Pomeron properties are dependent on the strong coupling boundary
conditions which are basically the frozen $\as$ and cut-off cases
considered in Sec.~\ref{s:cm}.

The first case is fairly simple, because of the fixed value of $\ab$
for $t<\tb$. In this region, because of the replacement
\begin{equation*}
 \exp\left\{A_1(\om)\int_{t'}^t\ab(\tau)\,\dif\tau\right\}\to
 \esp{\ab A_1(\om)(t-t')}
\end{equation*}
the solution is a plane wave whose momentum $\kappa$ is given in
Eq.~(\ref{kapa}). Therefore, $\op$ defines the boundary of the
continuum spectrum, given by $\kappa=0$, or
\begin{equation}
  \op = \frac{4\asb}{1 + \op - 2\asb A_1}\,.
\end{equation}
This is to be compared with the saddle point definition of the hard
Pomeron~\cite{CCS1} $\om_s(\tbar)$, which is obtained by
minimising the solution (as a function of $\ga$) of
\begin{equation}
  \om_s =\ab\left(\chi_0^{\om_s}(\ga) + \om_s
  \frac{\chi_1^{\om_s}}{\chi_0^{\om_s}}\right)\,.	\label{oms}
\end{equation}
Taking the $b=0$ form for $\chi_1^{\om}$ the minimum is at $\ga=1/2$,
and $\om_s$ is given by
\begin{equation}
  \om_s = \frac{4\asb}{1 + \om_s - 2\asb A_1}\,.
\end{equation}
which is identical to the true $\op$.

In \cite{CCS1} a second critical $\om$-exponent was proposed, termed
$\om_c$, corresponding to the position of the rightmost singularity of 
the anomalous dimension of the integrated gluon distribution. In the
collinear model the integrated gluon distribution can be defined as
\begin{equation}
  \label{eq:igluondef}
  g_\om(t)=\int_{-\infty}^t\dif t'\;\exp\left[-{1+\om\over2}(t-t')+
 A_1\int_{t'}^t\dif\tau\;\ab(\tau)\right]\cF(t')=
 \frac{1}{\asb(t)}\cFL(t)\,,
\end{equation}
because the corresponding density (at energy-scale $k^2$)
$\esp{{1+\om\over2}t}g_\om(t)$ can be shown to satisfy, in the collinear
limit, the usual DGLAP equation with anomalous dimension
$({1\over\om}+A_1)\ab$. The singularity of the effective anomalous
dimension
\begin{equation}
  \ga = \frac{\dif}{\dif t} \ln g_\om(t)
\end{equation}
is at the point $\om=\om_c(t)$ where $g_\om(t)$ goes to zero, i.e.,
where $\cFL(t)$ goes to zero.

This critical exponent $\om_c$ is related to $\op$ for the cutoff
case, in which $\as(t)=0$ for $t<\tb$. In fact, since
$\cFL(t)\sim\ab(t)$, we have $\cFL=0$ for $t<\tb$; but we noticed in
Sec.~\ref{s:cm} that $\dfrac{\cFL(t)}{\ab(t)}\sim g_\om(t)$ is instead
continuous at $t=\tb$, so that for $t\to\tb_+$, $\op$ satisfies the
boundary condition
\begin{equation}
 \cFL(\tb_+)=\ab(\tb_+){\cFH(\tb)\over\op(1+\op)}\neq0\,.\label{discA}
\end{equation}
This means that $\om_c(\tb)<\op$, because the depth of the well,
determined by $\om<\om_s(\tb)$, is to be further decreased below $\op$
in order to have $\cFL(\tb_+)=0$.

The relationships just found
($\op^{\rm freezing}=\om_s(\tb),\;\op^{\rm cutoff}>\om_c(\tb)$)
represent two extreme cases of the boundary condition dependence of
$\op$. If the strength $\ab(t)>0$ is positive but has intermediate
size and shape for $t<\tb$, we expect in general that
\begin{equation}
 \om_c(\tb)<\op<\om_s(\tb)\,,\label{teww}
\end{equation}
i.e., the lower and upper bounds mentioned in Ref.~\cite{CCS1}.

\subsection{High energy behavior and diffusion}\label{ss:heb}

It is widely believed that a two-scale process --- described by a
small-$x$ equation of BFKL type --- is perturbative for large enough
$t$ and $t_0$, while it becomes a strong coupling process if the
energy is so large as to allow diffusion to small values of $t\simeq0$
($k^2\simeq\La^2$) \cite{FRBook,BaLo93}.

In the collinear model the Green's function has the explicit
expression (\ref{fattg}), in which the strong-coupling information is
clearly embodied in the ``$S$-matrix coefficient'' $S(\om)$. Therefore 
it allows a direct study of the relative importance of the
``perturbative'' part $\cF_R\otimes\cF_I$ and of the ``strong-coupling''
part $S\,\cF_R\otimes\cF_R$, induced by diffusion through the boundary
conditions at $t=\tb$.

For large $t$ and $t_0$, $\GG_\om$ takes the approximate WKB form
(\ref{gwkb}), that we study in the special case $t=t_0$, so
that
\begin{equation}
 \GG(Y;t,t)\simeq\int\difo\;\esp{\om Y}{w(t)\over2\om^2\kw_\om(t)}
 \left[1+S(\om)\exp\left(-2\int_{t_s}^t\kw_\om(\tau)\,\dif\tau\right)\right]
 \,,	\label{gytt}
\end{equation}
provided $\int_{t_s}^t\kw_\om(\tau)\,\dif\tau\gg1$.

We now notice that if $Y=\log(s/kk_0)=\log s/\La^2-t$ is not too large, 
the expression (\ref{gytt}) is dominated by a saddle point at
$\om=\omb$, such that
\begin{equation}
 \left.Y=\frac12{\de\over\de\om}\log\kw^2\right|_{\omb}\simeq{bt\over
 2\left(b\omb t-\chi_m\right)}\,,	\label{Y}
\end{equation}
and therefore
\begin{equation}
 \omb(Y,t)\simeq\om_s(t)+{1\over2Y}={\chi_m\over bt}+{1\over2Y}\,,
\qquad\chi_m=4\left(1+\om_s-{2A_1-b\over bt}\right)^{-1}\,.\label{omb}
\end{equation}
so that $\omb(Y,t)$ is not much different from $\om_s(t)$, the saddle
point exponent mentioned before (Eq.~(\ref{oms})). Furthermore, at
this saddle point the phase function takes the value
($\chi_m''\simeq\chi_m^3/2$)
\begin{equation} 
 \int_{t_s}^t\kw(\tau)\,\dif\tau\simeq\sqrt{2\over\chi_m''}
 \int_{t_s}^t(b\omb\tau-\chi_m)^{\half}\,\dif\tau
 \simeq{1\over3\chi_m\sqrt{\chi_m''}}\left({bt\over Y}\right)^{3\over2}
 \gg1\,,	\label{phase}
\end{equation}
provided
\begin{equation}
 bt\sim\om_s^{-1}\ll2Y\ll\left({4\over3b}\right)^{2\over3}
 \om_s^{-{5\over3}}\sim\as^{-{5\over3}}\,.     \label{cphase}
\end{equation}

In other words, the saddle point (\ref{omb}) is self consistent, i.e., 
it exists in the WKB region of Eq.~(\ref{cphase}), provided the
effective parameter $\as^{5/3}Y\ll1$.

The actual evaluation of (\ref{gytt}) is now performed by distorting
the $\om$ contour as in Fig.~\ref{f:contour} and picking up the
saddle-point value of the background integral and the $\om$-poles:
\begin{equation} 
 \GG(Y;t,t)\simeq{1\over\sqrt{2\pi\chi_m''\ab Y}}\,\esp{\om_s(t)Y}\times
 \left[1+\ord(\as^5Y^3)\right]
 +\esp{\op Y}R_{\mathbb{P}}\esp{-(1+\op)t}\,.\label{res}
\end{equation}
\FIGURE{%
    \epsfig{file=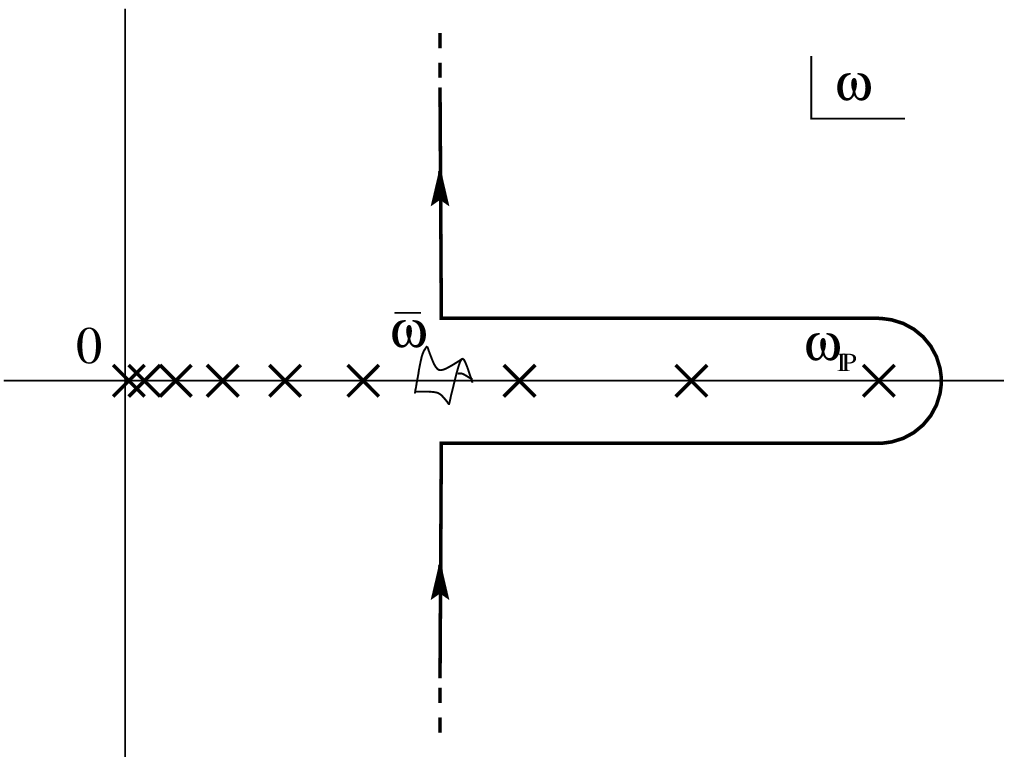,width=0.59\textwidth}
    \caption{The contour of integration for the Green's function
	$\GG_\om(Y;t,t)$.}
    \label{f:contour}}

Here we have kept the leading Pomeron pole, with residue
$R_{\mathbb{P}}$, and neglected the nearby poles
$\om_n\gtrsim\omb=\om_s+{1\over2Y}$, which are anyhow exponentially
suppressed because of the large phase function
$\ord\left((\as^5Y^3)^{-1/2}\right)$ of Eq.~(\ref{phase})\footnote{%
Of course, the background integral contribution is equivalent to the
one of the poles $\om_n\lesssim\omb$, as it can be seen by displacing
the contour further to the left.}.

The corrections to Eq.~(\ref{gytt}), depending on the (small)
parameter $\as^5Y^3$, can be explicitly evaluated from higher order
WKB terms. More simply, we can use the fact that the effective $\om$
values are close to the minimum of $\chi$, in order to use the Airy
form~\cite{CC96} of the Green's function, which in the cutoff case is given
by the expression
\begin{equation}
 \GG_\om(t,t_0)={t_0\over\om}\left({2b\om\over\chi_m''}\right)^{2\over3}
 \Ai(\xi)\left[\Bi(\xi_0)-{\Bi(\xib)\over\Ai(\xib)}\Ai(\xi_0)\right]
 \label{gairy}
\end{equation}
where the variables
\begin{equation}
 \xi=\left({2b\om\over\chi_m''}\right)^{1\over3}
 \left(t-{\chi_m\over b\om}\right)\,,
 \qquad\xib=-\left({2b\om\over\chi_m''}\right)^{1\over3}
\left({\chi_m\over b\om}
 -\tb\right)\,,	\label{xi}
\end{equation}
evaluated at $\om\simeq\omb$, turn out to be both large parameters
$\sim t^{2/3}$ of opposite sign. By using the expansion~\cite{abram}
\begin{equation}
 \Ai(\xi)\Bi(\xi)\simeq{1\over2\pi\sqrt{\xi}}\left(1+{c\over\xi^3}
 \right)\,,\qquad c={5\over32}	\label{AiBi}
\end{equation}
and the saddle-point value
\begin{equation*}
 \xi^{3\over2}(\omb,t)\simeq\sqrt{2\over\chi_m''}{t\over\chi_m}
 \left({bt\over2Y}\right)^{3\over2}\,,
\end{equation*}
we see that the correction is indeed of order $\ab^5Y^3$. The
quantitative evaluation requires a careful treatment of
$\om$-fluctuations around $\om-\om_s=1/2Y$. By using Eq.~(\ref{AiBi})
and the integral
\begin{equation*}
 \int_{\epsilon-\ui\infty}^{\epsilon+\ui\infty}{\dif x\over2\ui\sqrt{\pi x}}
 \;(2x)^{-3}\esp{x}={1\over15}\,,
\end{equation*}
we finally obtain, for the perturbative part of the Green's function
(\ref{res}) the sub-asymptotic correction factor
\begin{equation}
 1+{1\over24}(\chi_m^2\chi_m''b^2)\ab^5Y^3\,.	\label{subcor}
\end{equation}
The latter turns out to coincide with the ``non-Regge correction'' to the
``Regge exponent'' $\om_s(t)Y$ found by other
authors~\cite{Levin} in different but related contexts.

\!We notice, however, that the true Regge contribution is the second
term in Eq.~(\ref{res}), which is of strong-coupling type, with a
$t$-independent and eventually leading exponent $\op$. The
perturbative part, which dominates in the large-$t$ limit, comes from
the background integral and has no reason to be Regge behaved.

Thus, the appearance of the parameter $\ab^5Y^3$ is Eq.~(\ref{AiBi})
signals just the existence of a ``quantum'' wavelength in the solution
for $\om\lesssim\om_s(t)$. When $\ab^5Y^3\sim1$ the saddle point
breaks down and the solutions enter the small-$t$ regime.

It is non trivial, however, that in the well defined intermediate regime
(\ref{cphase}), the exponent $\om_s(t)$ with the corrections
(\ref{subcor}), appears to be an observable quantity.
On the other hand, the exponent $\om_c(t)$ --- the formal anomalous
dimension singularity --- does not directly appear in the $Y$
dependence, because the oscillating behavior of the $\cF$'s is masked
by the onset of $\op$ dominance for
\begin{equation}
 Y>Y_t\equiv\frac{1}{\op-\om_s(t)}t\,.	\label{Yt}
\end{equation}

We conclude from Eqs.~(\ref{gytt}) and (\ref{res}) that the two-scale
Green's function shows a perturbative (non-Regge) regime where the
exponent $\om_s(t)Y$ shows up with calculable corrections
(Eq.~(\ref{subcor})), provided the parameter $\ab^5Y^3$ is small. Even 
before the latter gets large, at $Y\gtrsim Y_t$ the Pomeron-dominated
regime takes over, characterized by the regular solution, which is
confined to the strong-coupling region of small $t$'s.

\section{Numerical results}

Here we concentrate on a couple of numerical aspects. Firstly, the
direct calculation in $t$-space of the regular solution: this allows
to test the $\ga$-representation method used for the full small-$x$
equation~\cite{CCS1}. Secondly, the direct evolution in $Y$ space of
the Green's function, in a simplified case where the $\om$ and $A_1$ 
dependences
are removed: this allows us to elucidate the transition region between the 
two asymptotic regimes just mentioned.

\subsection{Regular solution}
\label{sec:RegSoln}

\paragraph{General structure.} The regular solution $\cF_R$ of the
differential equation is obtained by starting at large $t$ with an
arbitrary initial condition and then evolving downwards. Since the
irregular solution falls rapidly relative to the regular solution when
reducing $t$, one quickly reaches a situation where only the regular
solution is left. By starting at large enough $t$ one can ensure that
this be true to arbitrary accuracy.

In \cite{CCS1} much use was made of the $\om$-representation to obtain
an approximation to the regular solution. In this section we wish to
carry out several tests of the practical accuracy of the
$\om$-representation.  All results shown here were obtained using
$A_1=-1$ and $b=1$.

First we illustrate the form of the unintegrated distribution in
Fig.~\ref{fig:freg}, normalised to be $1$ at the maximum.  Results
are shown both from a direct solution of the differential equation and
from the $\om$-representation. One sees that for larger values of $t$,
the $\om$-representation is in good agreement with the exact solution,
while for smaller $t$, where the solutions oscillate the results from
the $\om$-representation are slightly out of phase with the exact
solution. In general we are interested in the behavior to the right
of the rightmost zero.
\FIGURE{%
   \epsfig{file=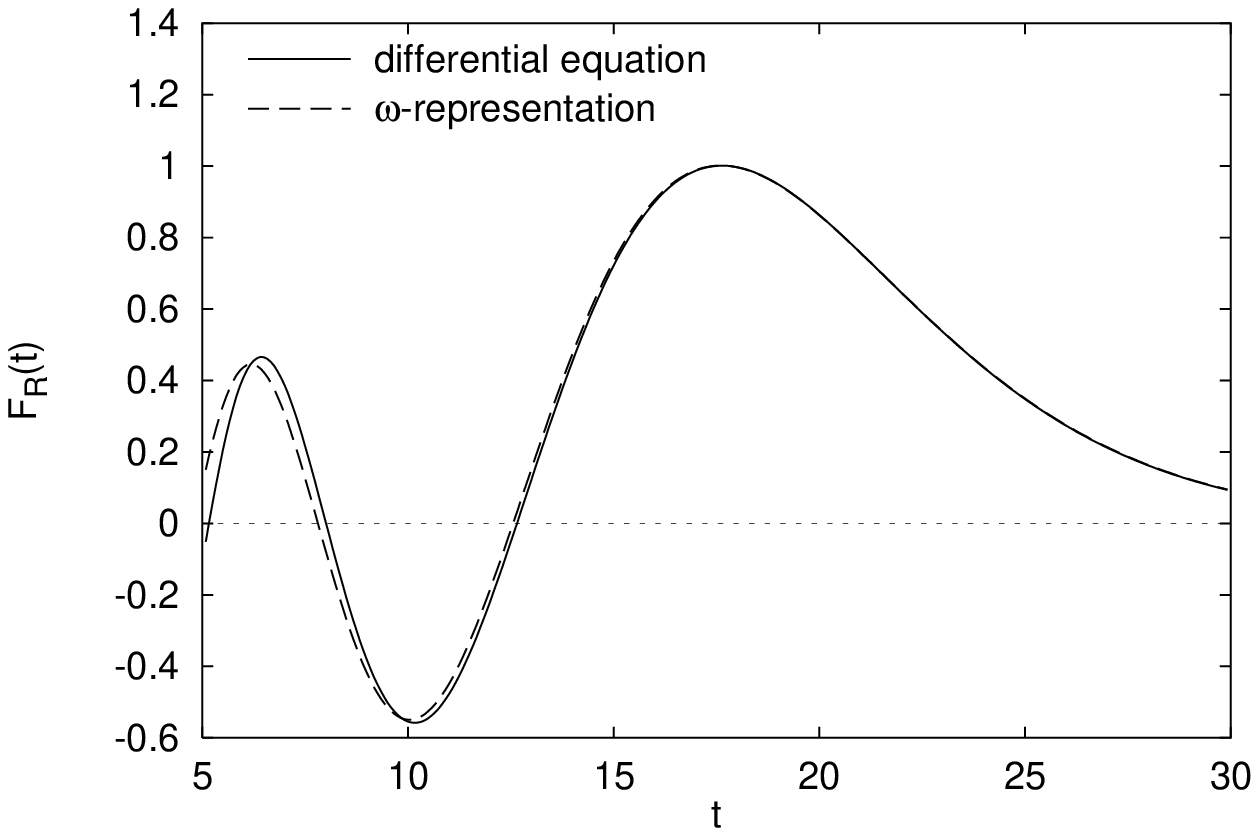,width=0.7\textwidth}
    \caption{Regular (unintegrated) solution from the explicit
      solution of the differential equation and from the $\om$
      representation; shown for $\om=0.15$. }
    \label{fig:freg}}

\paragraph{Critical exponent.} One quantitative test of the
$\om$-representation concerns the critical exponent $\om_c$, i.e.\ %
the value of $\om$ at which the anomalous dimension diverges. We
recall that this is connected with the position of the rightmost zero
of the regular solution: $\cF_{\om_c}(t)=0$. The $\om$-representation
determination of the position of this zero, or equivalently its
determination of $\om_c$ as a function of $t$, involves a small error
which we call $\delta\om_c$.  Fig.~\ref{fig:omtdiff} shows $\delta
\om_c/\om_c$ as a function of $\om_c$, and we see that the relative
error on $\om_c$ goes roughly as $\om^2$, or equivalently as $\asb^2$.
This corresponds to a NNL correction and is beyond our level of
approximation. We note also that even for relatively large values of
$\om\sim0.3$, the relative correction remains of the order of $5\%$
which is quite acceptable. In other words the NNL correction that
arises is not accompanied by a large coefficient.
\FIGURE{
    \epsfig{file=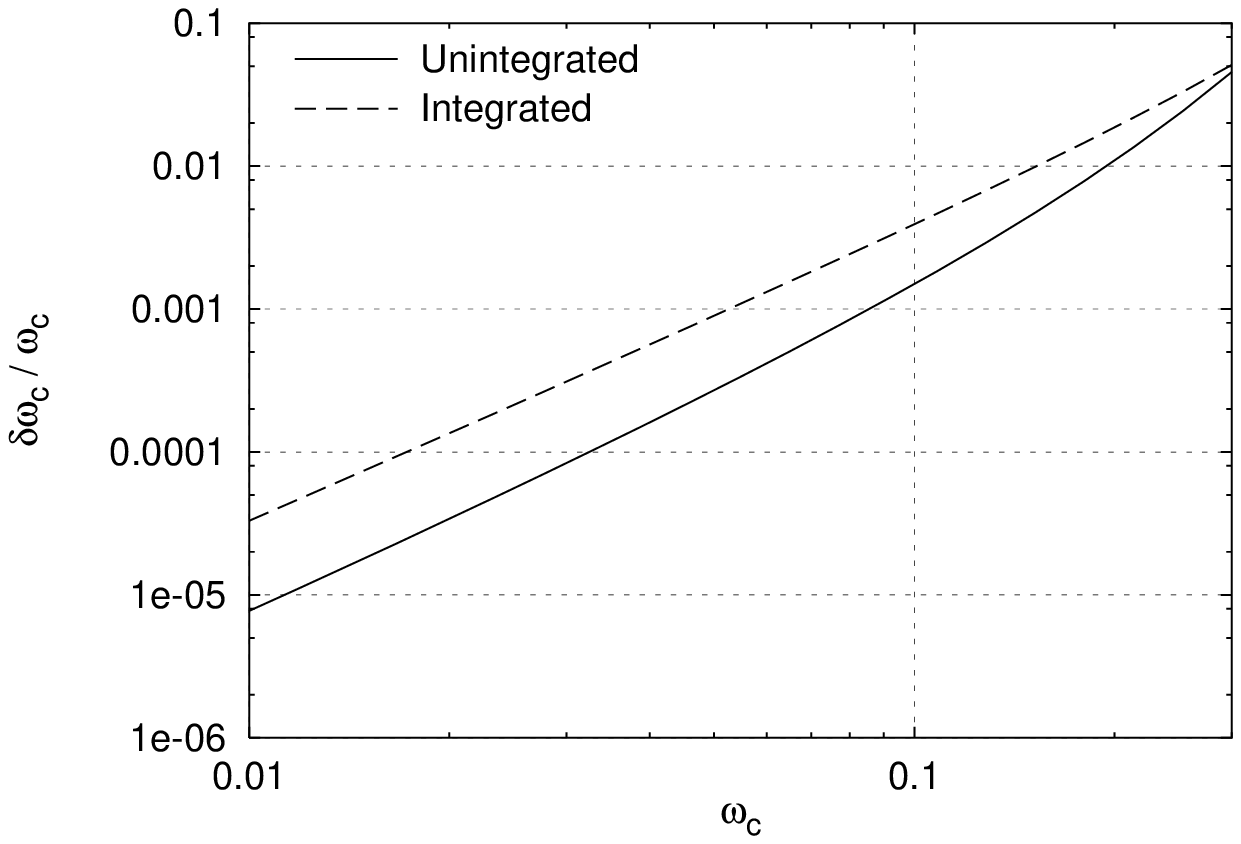,width=0.7\textwidth}
    \caption{The error, $\delta \om_c$ in the determination of $\om_c$
      within the $\om$ representation. Shown for both the unintegrated
      and integrated solutions.}
    \label{fig:omtdiff}}

\paragraph{Anomalous dimensions.} Our second quantitative test of the
$\om$-representation concerns the anomalous dimension. The error in
the $\om$-representation anomalous dimension, $\delta \ga$, is plotted
in Fig.~\ref{fig:dgam} as a function of $\asb$ for two values of
$\om$.  Let us first concentrate on the region for $\asb < 0.01$. We
see that the error is roughly independent of $\om$, and proportional
to $\asb^2$. In other words the difference between the exact result
and the $\om$-representation is a term of $\cO{\asb^2}$. We recall
that the terms that we wish to include properly are the leading terms
$(\asb/\om)^n$, the NL terms $\asb(\asb/\om)^n$ and the collinear
terms $(\asb/\om) \om^n$. A first correction of $\cO{\asb^2}$ is
consistent with all these terms having been correctly included.

For $\asb > 0.01$ we see that the error in the anomalous dimension has 
a more complicated behavior: it changes sign (the dip) and then
diverges, at the same point as the divergence in the anomalous
dimension itself (solid curve): this is just a consequence of the
position of the divergence of the anomalous dimension from the
$\om$-representation being slightly different from the true position
(cf.~Fig.~\ref{fig:omtdiff}).
\FIGURE{%
   \epsfig{file=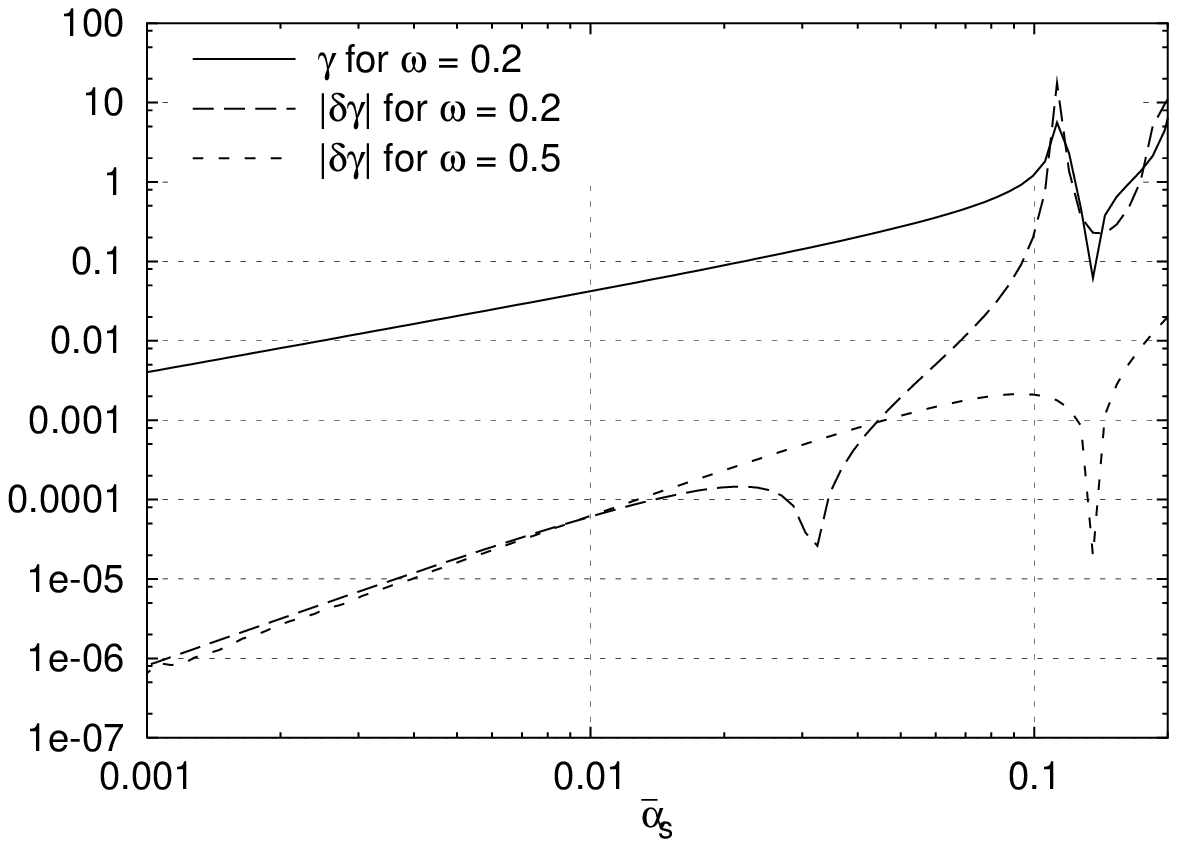,width=0.7\textwidth}
    \caption{Error in the anomalous dimension from the
	$\om$-representation.}
    \label{fig:dgam}}

\subsection{Green's function with running coupling, diffusion,
	the infra-red}\label{ss:infra}

We have already discussed in Sec.~\ref{ss:heb} the analytical
expression of the Green's function of our model, and the features of
the perturbative and strong-coupling regimes in the
$Y$-dependence. Here we want to study the transition region between
the two, which is analytically hard to describe.

Note first that several aspects of the regimes are due to just running
$\as$ effects. The latter enters at two levels: first, because of the
``acceleration'' in the well of Fig.~\ref{f:potenziale}, it modifies
the traditional fixed-$\as$ kind of diffusion, weighting it towards
lower transverse scales --- this may well be connected with the
breakdown of the saddle-point approximation for $\as^{5/3} Y \sim 1$
as discussed above. Secondly, because of the strong-coupling boundary
condition, it also introduces a qualitatively new kind of
diffusion, perhaps more properly referred to as `tunneling': namely
there is a certain $Y = Y_t$, defined in Eq.~(\ref{Yt}), at which the
non-perturbative pomeron suddenly takes over, because
\begin{equation}
  \esp{\op Y - t}\gtrsim\esp{\om_s(t)Y}	\label{opover}
\end{equation}
and beyond which the Green's function is dominated by the regular
solution, and thus confined to small $t$-values.

This `tunneling' phenomenon is qualitatively different from
diffusion, in so far as there is not a gradual decrease in the
relevant scale for the evolution, until non-perturbative scales become
important, but rather there is a point beyond which low scales
suddenly dominate, without intermediate scales ever having been
relevant. To see this consider that the contribution to the evolution
at scale $t$ from an intermediate scale $t_i \ll t$ (with a
corresponding exponent
$\om_i\simeq \chi_m/b t_i$), is of order
\begin{equation}
  e^{\chi_m Y/(b t_i) - (t-t_i)}\,.
\end{equation}
From this one sees, because of Eq.~(\ref{opover}), that $t_i=\tbar$
becomes relevant before higher scales do.

To study these effects at a qualitative level it suffices to consider
a very simplified version of the collinear model: one which retains
only the running of the coupling, but not the $\om$-shifts of the
$\ga=0,1$ poles, nor the $A_1$ component of the NL
corrections. We then examine the solution to 
\begin{equation}
  \GG(Y,t,t_0) = \delta(t-t_0) \delta(Y) + \int_0^Y\dif y\;K\otimes\GG(y)\,,
\end{equation}
for this simplified kernel and we consider the effective exponent of
the evolution
\begin{equation}
  \om_\eff = \frac{\dif}{\dif Y} \ln\GG(Y,t,t)
\end{equation}
as a function of $Y$. Fig.~\ref{fig:expstep} illustrates the basic
behavior of $\om_\eff$ for rather extreme kinematics --- not intended
to be phenomenologically relevant, but rather to show clearly the
relevant features. Two values of the infra-red cutoff are
considered. What is seen is that the exponent at first increases
slowly and smoothly, and then at a certain threshold $Y$ increases
rapidly towards $\op$. The saturation of $\op$ occurs later in $Y$ for
increased $\tbar$ (i.e.\ decreased $\op$) as expected from
\eqref{Yt}. Traditional smooth diffusion into the infra-red would have
led to the opposite behavior, namely the higher $\tbar$ case (lower
$\op$) being saturated first.
\FIGURE{
   \epsfig{file=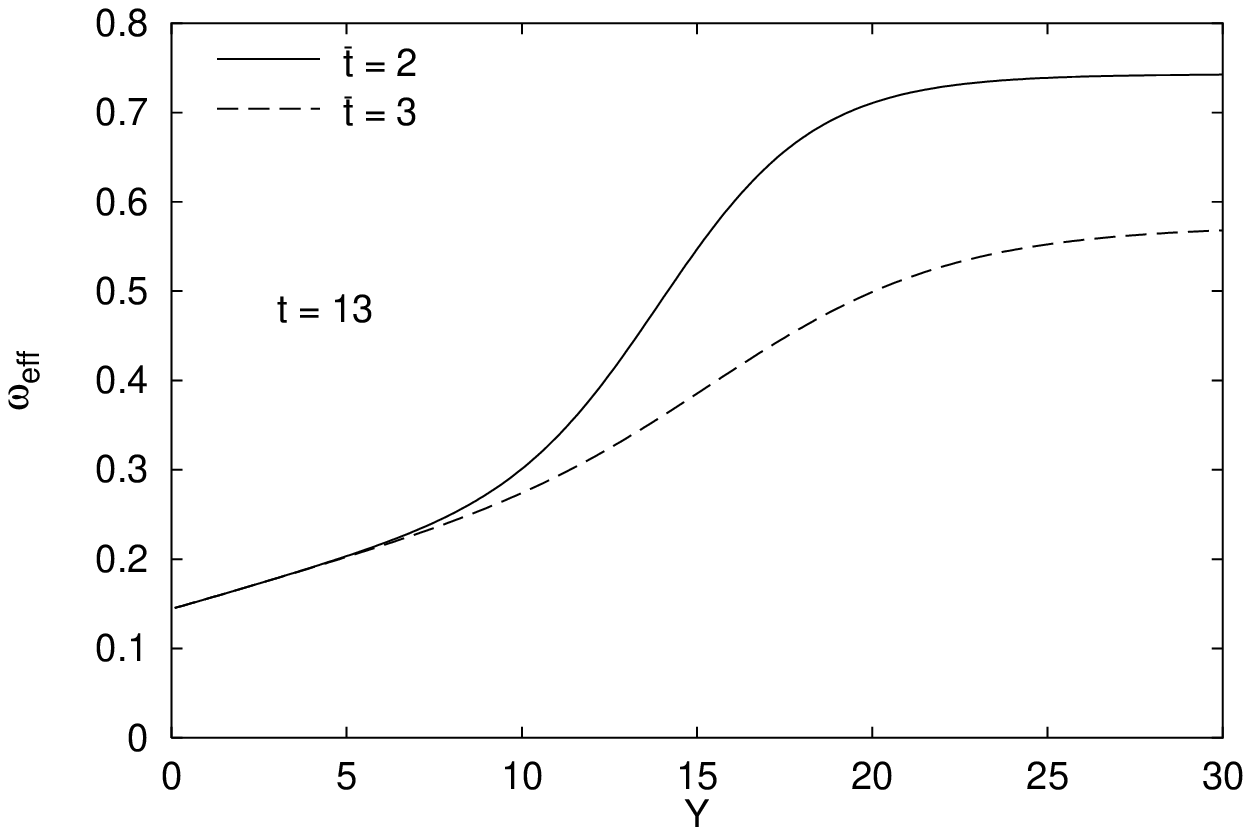,width=0.7\textwidth}
    \caption{The effective intercept $\om_\eff$ as a function of $Y$ for
      $t=13$ and two different values of the infra-red cutoff
      $\tbar$.} 
    \label{fig:expstep}}

The study of traditional diffusion is a little trickier, essentially
because for most parameter choices there is only a limited domain of
traditional diffusion before tunneling takes over. Nevertheless one
way to approach it is to study the difference between the effective
exponents in a case with running $\as$ and one with fixed $\as$. For
this comparison to be meaningful the fixed $\as$ must be chosen such
that $\as \chi_0(1/2) = \om_s(t)$. The observed difference in exponent 
is then a measure of the difference in the diffusion properties of the 
running and fixed-$\as$ cases: it is sensitive to the terms which are
expected to be sizeable when $\as^5 Y^3\sim 1$. 
\FIGURE{%
    \epsfig{file=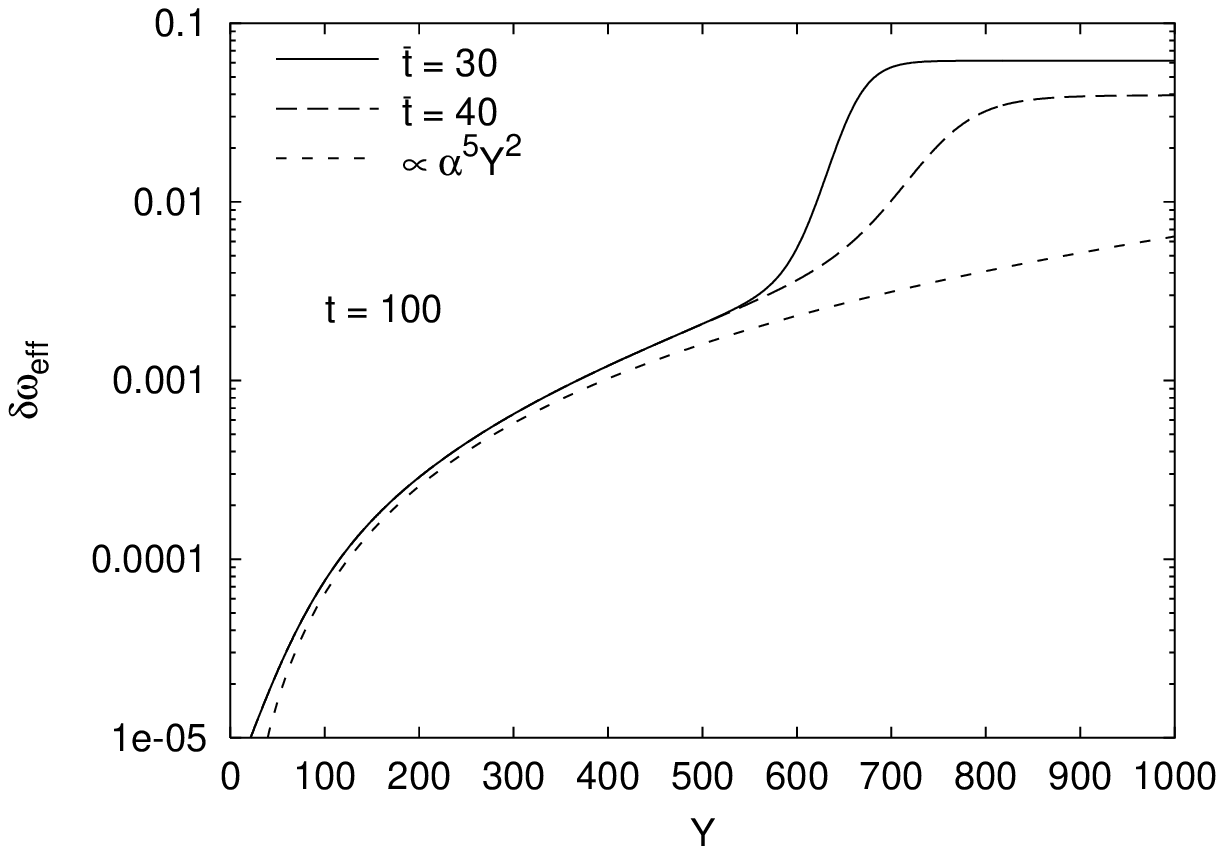,width=0.7\textwidth}
    \caption{The difference $\delta \om_\eff$ between the effective
      exponents seen with running $\asb$ and fixed $\asb$ ($=1/101$),
      shown for two values of $\tbar$. Shown for very extreme values
      of $t$ and $Y$ in order to expose more clearly the main features.}
    \label{fig:rfdiff}}

The difference is shown in Fig.~\ref{fig:rfdiff}, where the dashed
line is based on Eq.~(\ref{subcor}), which in turn reproduces the
results by Kovchegov and Mueller and by Levin~\cite{Levin} 
(with the proper changes for the $\chi_m$ and $\chi_m''$ values). The
agreement is good, up to the critical value $Y_t$ beyond which the
Pomeron regime takes over.

In Fig.~\ref{fig:rfdiff} we have taken extreme values of $t$ and $Y$,
in order to identify the running coupling corrections, and we have
thus emphasized the difference $\op-\om_s(t)$ also. With realistic
parameters and model, the transition between the two regimes could be
smoother and shifted to relatively higher energies.

\section{Discussion}

The model we have introduced is somewhat intermediate between the full 
RG-improved small-$x$ equation and the (Airy) diffusion model with
running coupling. Compared to the latter, it has the advantage of a
correct collinear behavior, still admitting a Schr\"odinger-type
treatment. It appears, therefore, as a useful laboratory for testing
approximation methods and theoretical prejudices.

We have already performed several tests. First, we have shown
(Sec.~\ref{sec:RegSoln}) that indeed the $\om$-expansion of
the $\ga$-representation provides a quite good approximation of the
regular solution, which is thus perturbative, i.e., independent of the
boundary conditions in the strong-coupling region. This lends
further support to previous results~\cite{CCS1} on the anomalous
dimensions and the hard Pomeron.

We have further used the present model to analyze the gluon Green's
function, by testing its factorisation properties, and by providing
its explicit expression (Sec.~\ref{s:af}) which involves the
left-regular solution, and is thus dependent on the strong-coupling
region. In the present model, this dependence occurs through a single
scattering coefficient $S(\om)$, which carries the spectrum and thus
the leading Pomeron singularity (Eq.~(\ref{fattg})). It is expected,
by the arguments of Ref.~\cite{CC96}, that a similar decomposition may 
be valid in general, perhaps with several scattering coefficients.

For two-scale processes, it is of particular interest the result of
Secs.~\ref{ss:heb} and \ref{ss:infra}: in an intermediate $Y$ region a 
perturbative regime exists where the exponent $\om_s(t)$ is
observable, with corrections of type $\ab^5Y^3$, whose size is in
agreement with Refs.~\cite{Levin}. On the other hand, for large
enough energies, $Y\gtrsim t/(\op-\om_s(t))$, the Regge behavior due
to the non-perturbative Pomeron takes over with a ``tunneling''
transition to small $t$'s which may be, in principle, spectacular for
sizeable values of $\op-\om_s(t)$.

Several points need further elucidation. First, a realistic evaluation 
of $\op$ and of the related effective coupling is still needed, in
order to put the question of the regimes in perspective, and to start
making predictions for two-scale processes. In particular, the size of
$\op$ (including possible unitarity corrections) seems to be crucial
in order to understand how fast and experimentally relevant is the
transition to the Pomeron regime.

Furthermore, the relative importance of the perturbative exponents
$\om_s(t)$ and $\om_c(t)$ and of $\op$ is still to be cleared up in
DIS-type processes, where $t$-evolution plays a much more important
role.

For all these questions the present model is likely to provide useful
hints and a preliminary understanding.

\section*{Acknowledgements}

One of us (G.P.S.) wishes to acknowledge stimulating conversations with Raju
Venugopalan on issues related to diffusion and running coupling.


\begin{thebibliography}{99}
\bibitem{BFKL} L.N. Lipatov, \sjnp{23}{1976}{338};\\
       E.A. Kuraev, L.N. Lipatov and V.S. Fadin, \jetp{44}{1976}{443};\\
       E.A. Kuraev, L.N. Lipatov and V.S. Fadin, \jetp{45}{1977}{199};\\
       Ya. Balitskii and L.N. Lipatov, \sjnp{28}{1978}{822}.
\bibitem{NLL}
  L.N. Lipatov and V.S. Fadin, \sjnp{50}{1989}{712};\\
  V.S. Fadin, R. Fiore and M.I. Kotsky, \plb{359}{1995}{181};\\ 
  V.S. Fadin, R. Fiore and M.I. Kotsky, \plb{387}{1996}{593}
	[\hepph{9605357}];\\ 
  V.S. Fadin, and L.N. Lipatov, \npb{406}{1993}{259};\\
  V.S. Fadin, R. Fiore and A. Quartarolo, \prd{50}{1994}{5893}
	[\hepth{9405127}];\\
  V.S. Fadin, R. Fiore, and M. I. Kotsky, \plb{389}{1996}{737}
	[\hepph{9608229}];\\
  V.S. Fadin and L.N. Lipatov, \npb{477}{1996}{767} [\hepph{9602287}];\\
  V.S. Fadin, M.I. Kotsky and L.N. Lipatov, \plb{415}{1997}{97};\\
  S. Catani, M. Ciafaloni and F.Hautman, \plb{242}{1990}{97};\\
  S. Catani, M. Ciafaloni and F.Hautman, \npb{366}{1991}{135};\\ 
  V. S. Fadin, R. Fiore, A. Flachi, and M. I. Kotsky, \plb{422}{1998}{287}
	[\hepph{9711427}].
\bibitem{FL}
  V.S. Fadin and L.N. Lipatov, \plb{429}{1998}{127} [\hepph{9802290}].
\bibitem{CC98}
  M. Ciafaloni and G. Camici, \plb{412}{1997}{396} [\hepph{9707390}];\\
  M. Ciafaloni, \plb{429}{1998}{363} [\hepph{9801322}];\\
  M. Ciafaloni and G. Camici, \plb{430}{1998}{349} [\hepph{9803389}]. 
\bibitem{BV} J. Bl\"umlein, W.L. van Neerven, V. Ravindran and A. Vogt,
	\hepph{9806368};\\
  J. Bl\"umlein, W.L. van Neerven and V. Ravindran, \prd{58}{1998}{091502}
	[\hepph{9806357}].
\bibitem{Ross} D.A. Ross, \plb{431}{1998}{161} [\hepph{9804332}].
\bibitem{Levin} Yu.V. Kovchegov and A.H. Mueller, \plb{439}{1998}{428}
	[\hepph{9805208}];\\
	E. Levin, \hepph{9806228}.
\bibitem{Sa98} G.P. Salam, \jhep{9807}{1998}{19} [\hepph{9806482}].
\bibitem{CiCo98b}
  M. Ciafaloni and D. Colferai, \plb{452}{1999}{372} [\hepph{9812366}].
\bibitem{Th99} R.S. Thorne, \hepph{9901331}.
\bibitem{CCS1} M. Ciafaloni, D. Colferai and G.P. Salam, \hepph{9905566}. 
\bibitem{hera}  H1 Collaboration (C. Adloff et al.),\zpc{72}{1996}{593};\\
             ZEUS Collaboration (J. Breitweg et al.),\plb{407}{1997}{402}.
\bibitem{2dis}  L3 Collaboration (M. Acciarri et al.),\plb{453}{1999}{333}.
\bibitem{fjet} ZEUS Collaboration (J. Breitweg et al.),\epj{6}{1999}{239};\\ 
     H1 Collaboration (C. Adloff et al.),\npb{538}{1999}{3} [\hepex{9809028}].
\bibitem{DGLAP} V.N. Gribov and L.N. Lipatov, \sjnp{15}{1972}{438};\\
	G. Altarelli and G. Parisi, \npb{126}{1977}{298};\\
	Yu.L. Dokshitzer, \jetp{46}{1977}{641}. 
\bibitem{CC96}
  G. Camici and M. Ciafaloni, \plb{395}{1997}{118} [\hepph{9612235}].
\bibitem{GLR}
    L.V. Gribov, E.M. Levin and M.G. Ryskin,\prep{100}{1983}{1}.
\bibitem{CoKw89} J. Kwiecinski, \zpc{29}{1985}{561};\\
                      J.C. Collins and J. Kwiecinski, \npb{316}{1989}{307}.
\bibitem{BaLo93} J. Bartels and H. Lotter, \plb{309}{1993}{400}.
\bibitem{FRBook} see, e.g., J.R. Forshaw and D.A. Ross, {\em Quantum
Chromodynamics and the Pomeron}, Cambridge University Press, 1997.
\bibitem{abram} M. Abramowitz and I. Stegun,
	{\it Handbook of Mathematical Functions}, Dover Publication.

\end{thebibliography}
\end{document}